\documentclass[runningheads]{llncs}
\usepackage{graphicx}
\usepackage{amsmath}
\usepackage{amsfonts}
\usepackage{multirow} 
\usepackage[misc]{ifsym} 
\usepackage{soul,color}
\begin{document}
	\title{Realistic Ultrasound Image Synthesis for Improved Classification of Liver Disease}
	%
	\titlerunning{Realistic liver Ultrasound Image Synthesis}
	%
	
	\author{Hui Che\inst{1} \and
		Sumana Ramanathan\inst{1} \and 
		David J. Foran\inst{4} \and 
		John L. Nosher\inst{2} \and
		Vishal M. Patel \inst{3} \and
		Ilker Hacihaliloglu	\orcidID{0000-0003-3232-8193}\inst{1,2}}

\institute{Department of Biomedical Engineering, Rutgers University, NJ, USA
	\and 
	Department of Radiology, Rutgers Robert Wood Johnson Medical School, NJ, USA
	\and
	Department of Electrical and Computer Engineering, Johns Hopkins University, MD, USA
	\and
	Rutgers Cancer Institute of New Jersey, NJ, USA}

	\authorrunning{Che et al.}

\maketitle              
\begin{abstract}
With the success of deep learning-based methods applied in medical image analysis, convolutional neural networks (CNNs) have been investigated for classifying liver disease from ultrasound (US) data. However, the scarcity of available large-scale labeled US data has hindered the success of CNNs for classifying liver disease from US data. In this work, we propose a novel generative adversarial network (GAN) architecture for realistic diseased and healthy liver US image synthesis. We adopt the concept of stacking to synthesize realistic liver US data.  Quantitative and qualitative evaluation is performed on 550 in-vivo B-mode  liver US images collected from 55 subjects. We also show that the synthesized images, together with real in vivo data, can be used to significantly improve the performance of traditional CNN architectures for Nonalcoholic fatty liver disease (NAFLD) classification. 

\end{abstract}
\section{Introduction}
Nonalcoholic fatty liver disease (NAFLD) is being recognized as one of the most common liver diseases worldwide, affecting up to 30\% of the adult population in the Western countries \cite{targher}. It is defined as a condition with increased fat deposition in the hepatic cells due to obesity and diabetes in the absence of alcohol consumption \cite{deepak}. Patients with NAFLD are at an increased risk for the development of cirrhosis and hepatocellular carcinoma (HCC) which is one of the fastest-growing causes of death in the United States and poses a significant economic burden on healthcare  \cite{nasr2018natural}. Therefore, early diagnosis of NAFLD is important for improved management and prevention of HCC. Liver biopsy is considered the gold standard for diagnosing NAFLD \cite{gaidos}. However, biopsy is an invasive and expensive procedure associated with serious complications making it impractical as a diagnostic tool \cite{tapper2017use}. Incorrect staging in 20\% of the patients has also been reported due to sampling error and/or inter-observer variability \cite{tapper2017use}. Diagnostic imaging, based on Ultrasound (US), Magnetic resonance imaging (MRI), and Computed Tomography (CT), has been utilized as a safe alternative. Due to being cost-effective, safe, and able to provide real-time bedside imaging US has been preferred over MRI and CT \cite{li}. Nonetheless, studies have shown that the specificity and sensitivity of US to detect the presence of steatosis is very poor \cite{khov}. Furthermore, the appearance of the tissue can be very easily affected by machine acquisition settings and the experience of the clinicians \cite{rajendra,strauss}.

To overcome the drawbacks of US imaging and improve clinical management of liver disease, Computer-Aided Diagnostic (CAD) systems have been developed. With the success of deep learning methods in the analysis of medical images, recent focus has been on the incorporation of convolutional neural networks (CNNs) into CAD systems to improve the sensitivity and specificity of US in diagnosing liver disease \cite{biswas,byra2018transfer,che2021liver,liu,meng,reddy}. Although successful results were reported, one of the biggest obstacles hindering the improved adaptation of deep learning-based CAD systems in clinical practice is the unavailability of large-scale annotated datasets. The collection of large-scale annotated medical data is a very expensive and long process. Albeit there are many publicly available datasets and world challenges, the data available is still limited and the focus has been on certain clinical applications based on MRI, CT, and X-ray imaging. One of the most commonly used practices to overcome the scarce data problem is data augmentation based on image geometric transformation techniques such as rotation, translation, and intensity transformations \cite{ali2019data,che2021liver,milletari2016vnet}. However, these transformations result in images with similar feature distributions and do not increase the diversity of the dataset required to improve the performance of any CNN model. A new type of data augmentation is image synthesis using generative adversarial networks (GANs). GAN-based medical image synthesis has become popular for improving the dataset size and has been extensively investigated for improving classification and segmentation tasks \cite{kazeminia2020gans,yi2019generative}. However, to the best of our knowledge GAN-based image synthesis in the context of liver disease classification from US data has not been investigated previously.

In this work, a novel GAN-based deep learning method is proposed to synthesize B-mode liver US data. Our contributions include: 1- We propose a stacked GAN architecture for realistic liver US image synthesis. 2- Using ablation studies we show how the performance of state-of-the-art GAN architectures can be improved using the proposed stacked GAN architecture. Qualitative and quantitative evaluations are performed on 550 B-mode in-vivo liver US images collected from 55 subjects. Extensive experiments demonstrate our method improves traditional single-stage GANs to generate B-mode liver US data. We also show that by using a larger balanced NAFLD dataset, including real and synthesized data, the performance of the liver disease classification task can be improved by 4.34\%.  

\section{Materials and Methods}
\begin{figure}[h]
	\centering
	\includegraphics[width=8cm]{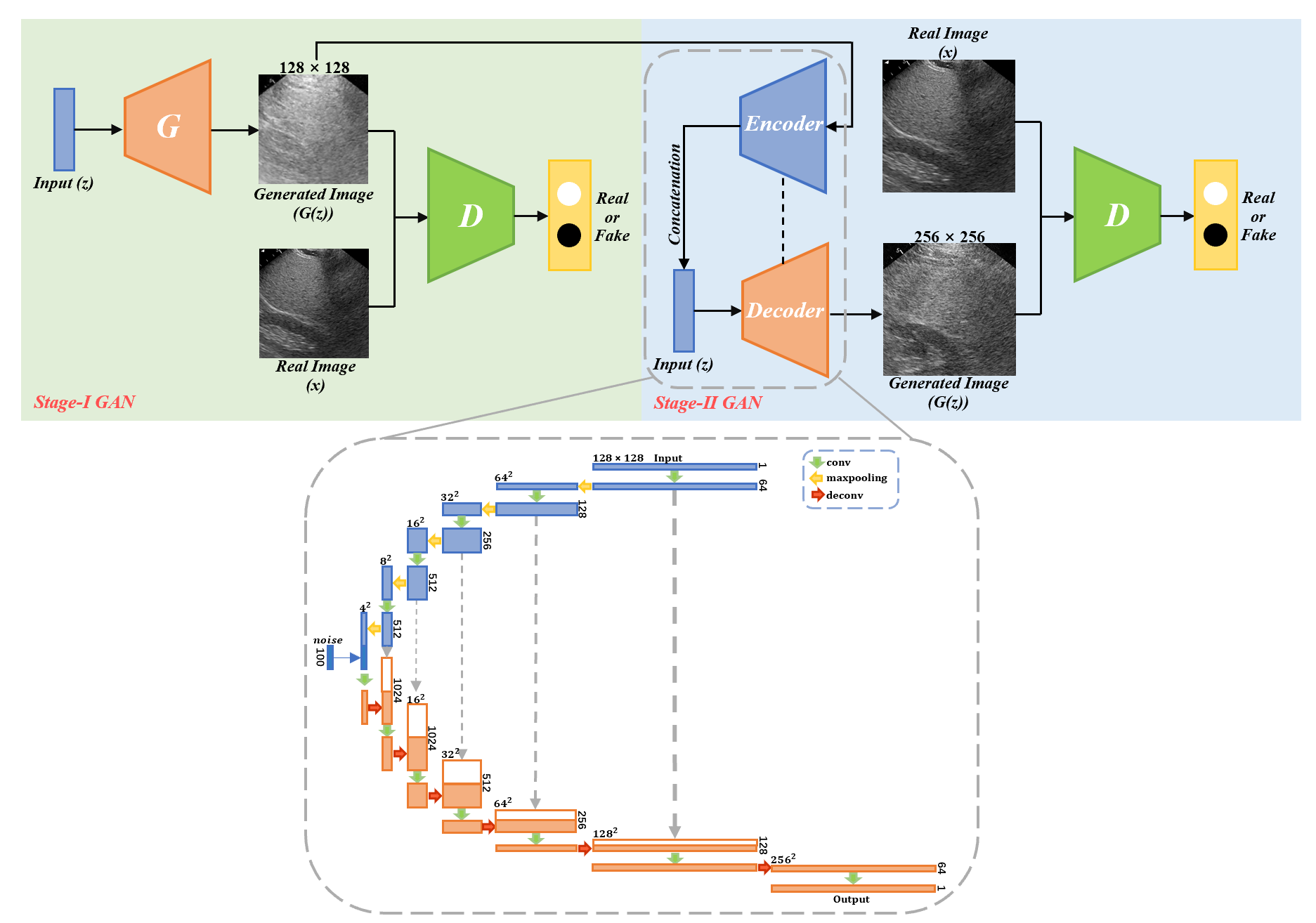}
	\caption{Overview of the proposed GAN-based network architecture. Stage-\uppercase\expandafter{\romannumeral1} GAN produces mid-resolution images and Stage-\uppercase\expandafter{\romannumeral2} GAN outputs high-resolution images with realistic tissue details. Dashed vertical lines represent skip-connection. Bottom: encoder-decoder architecture to integrate features.}
	\label{fig:fig1}
\end{figure}

The architecture of our proposed network is shown in Figure~\ref{fig:fig1}. Specifically we design a stacked GAN model (StackGAN)to generate high-resolution liver US images in two stages. A common GAN layout is utilized in Stage-\uppercase\expandafter{\romannumeral1} to synthesize a mid-resolution image. Stage-\uppercase\expandafter{\romannumeral2} GAN, the main contribution of this work, aims to output an image with improved tissue details by integrating features from the mid-resolution image during the generative process. The two stage approach overcomes the training instability of single stage GANs and results in improved and realistic representation of the synthesized liver US data. The overall network produces high-resolution images using random noise as input, not relying on any prior information from the original data.

\subsection{Stage-\uppercase\expandafter{\romannumeral1} GAN}
GAN \cite{goodfellow2020generative}, as an unsupervised generative model to learn the data distribution, is made of two distinct models: a generator \emph{G} to generate samples as realistic as possible and a discriminator \emph{D} to discriminate the belonging of the given sample. \emph{G} aims to transform a latent space vector $z \sim p(z) $ sampled from a prior distribution into a real-like image, while \emph{D} learns to distinguish between the real image $x$ and the fake image $G(z)$. GAN is trained by minimizing the following adversarial loss in an alternating manner \cite{zhang2019self}, which falls in a state of the confrontational game:

\begin{equation}
	\begin{aligned}
		&L_{D}=-\mathbb E_{x\sim p_{data}(x)}[logD(x)]-\mathbb E_{z\sim p_{z}(z)}[log(1-D(G(Z)))], \\
		&L_{G}=-\mathbb E_{z\sim p_{z}(z)}[logD(G(Z))]
	\end{aligned}
\end{equation}

In this work, for Stage-\uppercase\expandafter{\romannumeral1} GAN we adopt several popular GAN-based architectures to synthesize NAFLD US images: DCGAN \cite{radford2015unsupervised}, DCGAN with spectral normalization (SN) \cite{zhang2019self}, and DCGAN with SN \cite{miyato2018spectral} and self-attention module (SA) \cite{zhang2019self}. \\
\noindent\textbf{DCGAN:} DCGAN\cite{radford2015unsupervised} introduces CNN into the generative model to acquire the powerful feature extraction capability. Compared to the traditional GAN design, both the discriminator and the generator in DCGAN discard pooling layers and choose to use convolutional and convolutional-transpose layers respectively. 
The LeakyReLU activation is utilized in all layers of the discriminator to prevent gradient sparseness and the output of the last convolutional layer is processed by the Sigmoid function neither fully connected layer to give a discriminative result. In the generator, the output layer uses the activation function of $Tanh$ and the remaining layers use ReLU activations. Batch normalization in networks (not include the output layer of the generator and the input layer of the discriminator) helps prevent training issues caused by poor initialization. For uncoditional image synthesis, DCGAN has been widely adopted in the medical imaging community \cite{yi2019generative} and has therefore been chosen as our baseline GAN architecture. However, in the following sections we also introduce further improvements to DCGAN. 

\noindent\textbf{Spectral Normalization (SN) Module:} The main idea of SN is to restrict the output of each layer through the Lipschitz constant without complicated parameter adjustments \cite{miyato2018spectral}. Doing so can constrain the update of the discriminator triggered by generator update to a lesser extent. The normalization is applied in the generator and discriminator simultaneously. Most recently SN was incorporated into GAN for improving low dose chest X-ray image resolution \cite{xu2020low} and multi-modal neuroimage synthesis \cite{lan2020sc}. 

\noindent\textbf{Self-Attention (SA) Module:} SA module calculates the attention value between local pixel regions and helps to model global correlation in a wider range. The generator with the SA module learns specific structure and geometric features \cite{zhang2019self}. In addition, the discriminator can now perform complex geometric constraints more accurately on the global structure. This module was recently incorporated into a GAN architecture for synthesizing bone US data \cite{alsinan2020gan}.

\subsection{Stage-\uppercase\expandafter{\romannumeral2} GAN}
Directly generating high-resolution images usually meets problems of detail  and poor diversity. Instead, we turn to generate a mid-resolution image with high quality in Stage-\uppercase\expandafter{\romannumeral1} and obtain its feature maps in different depth, which are fused into corresponding layers of the generator for information supplement. The generated mid-resolution images have relatively diverse feature distributions but lack vivid tissue representations. The latter generator is supported by rich structure information from synthetic images with the size of $256\times256$ and fills feature details at the same time. 

The generated image in Stage-\uppercase\expandafter{\romannumeral1} GAN is fed into the Stage-\uppercase\expandafter{\romannumeral2} GAN generator. The Stage-\uppercase\expandafter{\romannumeral2} GAN generator is constructed using an encoder and decoder network architecture (Fig.\ref{fig:fig1}). The encoder receives Stage-\uppercase\expandafter{\romannumeral1} images and outputs feature maps in various sizes. The basic block in the encoder is comprised of a convolutional layer and a maxpooling layer. The downsampling enlarges the receptive field area and concentrates on feature extraction. Captured features are integrated into the generator by skip-connection. We also concatenate the random noise vector $z$ in the encoder output and input this combined feature vector to the decoder. Conditioned on the low-resolution result, obtained Stage-\uppercase\expandafter{\romannumeral1}, and the noise vector the discriminator and generator of Stage-\uppercase\expandafter{\romannumeral2} GAN are trained by maximizing $L_{D}$ and minimizing $L_{G}$ showed in Equation 1. The proposed method avoids the prior knowledge from real data as input and guarantees the diversity of generated images. Our discriminator, denoted as $D$, during this stage uses the same architecture of discriminator in DCGAN to perform differentiation of real or synthetic. Using information from generated mid-resolution images rather than real images prevents the model from memorizing patterns from real images. Furthermore, Stage-\uppercase\expandafter{\romannumeral2} GAN corrects imaging artifacts in the low-resolution image, obtained in Stage-I, synthesizing high-resolution realistic liver US data.

\section{Experiments and Results}
\subsection{Dataset}
Experiments are performed on the dataset provided by \cite{byra2018transfer}. The NAFLD dataset includes 550 B-mode US scans and biopsy results from 55 subjects. 10 US images were collected for each subject. Using biopsy, 38 subjects were diagnosed as NAFLD patients and the rest 17 were viewed as normal/healthy individuals. The data was collected using the GE Vivid E9 Ultrasound System (GE Healthcare INC, Horten, Norway) equipped with a sector US transducer operating at 2.5 MHz\cite{byra2018transfer}.  All images were cropped to remove irrelevant regions (mostly related to text involving image acquisition settings) and then resized to a size of $256\times256$. During the cropping, the original image resolution of 0.373 mm was kept constant.

\noindent\textbf{Training and test data:} To validate the performance of our proposed method, we randomly split 70 normal and 70 diseased images as a testing set, remaining images (100 healthy and 310 diseased) are grouped as a training set to train GAN-based networks. The classification network was trained using real and synthesized data totaling to 1000 healthy and diseased US images. The split obeys the rule that the same patient scans are not used for both training and testing. The random split operation was repeated five time and average results are reported. 

We conduct experiments using the PyTorch framework with an Intel Core CPU at 3.70 GHz and an Nvidia GeForce GTX 1080Ti GPU. GAN-based networks are trained using the cross-entropy loss and ADAM optimization method with batch size of 16 and a learning rate of 0.0002. Exponential decay rate for the first and second moment estimates are set to 0.5 and 0.999. The performance of our proposed generative method is compared with those popular GAN-based architectures mentioned in section 2.1 to generate liver US images directly. Examples for each class are generated individually, not incorporating class conditions.

\subsection{GAN-based Network Evaluation}
 \vspace{-0.5cm}
\begin{figure}[h]
	\centering
	\includegraphics[width=10cm]{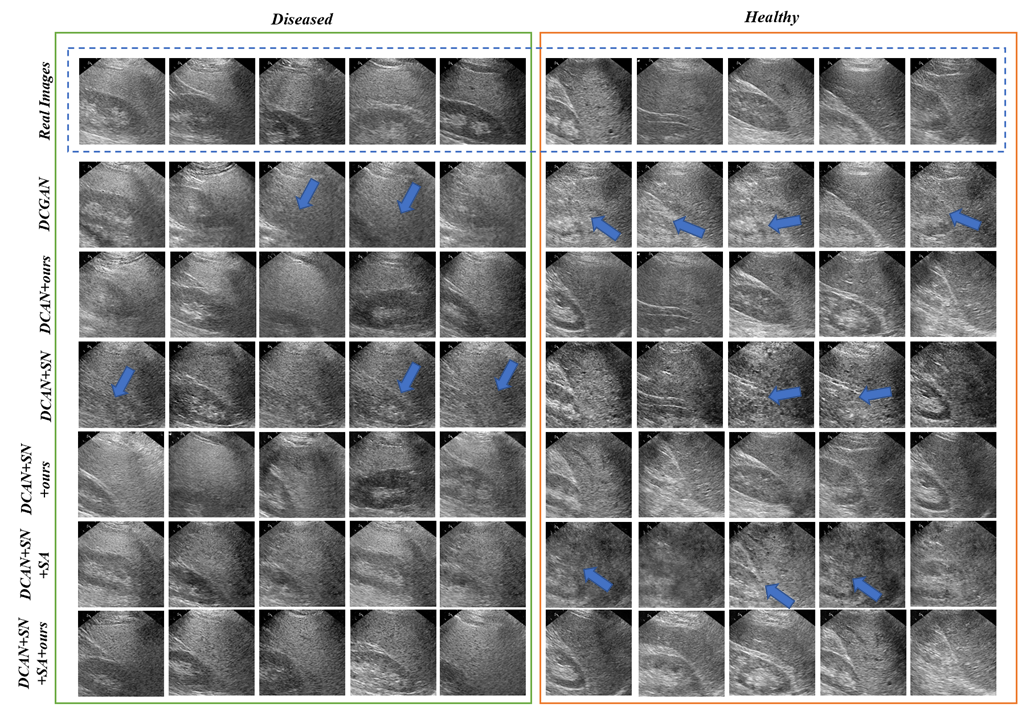}
		\caption{Qualitative results of the images generated by DCGAN in combination with the different modules, along with the proposed module. SA - Self-attention module and SN - Spectral normalization module. Blue arrows point to qualitative improvements achieved over prior state of the art. In all the presented results \textit{ours} denotes the integration of Stage-\uppercase\expandafter{\romannumeral2} GAN module.}
	\label{fig:fig2}
\end{figure}

\noindent\textbf{Qualitative Results:} Qualitative results for the investigated models are given in Fig.~\ref{fig:fig2}. The first row in the figure demonstrates examples of real images followed by the synthesized images generated by different methods, for both classes diseased and healthy. The different state of the art methods that are used to synthesize the images and compare them are, DCGAN\cite{radford2015unsupervised}, DCGAN combined with the proposed (Stage-\uppercase\expandafter{\romannumeral2} GAN) module, DCGAN\cite{radford2015unsupervised} with SN\cite{miyato2018spectral} module, DCGAN\cite{radford2015unsupervised} with SN\cite{miyato2018spectral} module combined with the proposed (Stage-\uppercase\expandafter{\romannumeral2} GAN) module, DCGAN\cite{radford2015unsupervised} with SN\cite{miyato2018spectral} and SA\cite{zhang2019self} modules, and a combination of DCGAN\cite{radford2015unsupervised} with all modules (SN\cite{miyato2018spectral}, SA\cite{zhang2019self} and proposed Stage-\uppercase\expandafter{\romannumeral2} GAN)). Investigating the results in Figure~\ref{fig:fig2} it can be seen that when the proposed module (Stage-\uppercase\expandafter{\romannumeral2} GAN) is used in combination with state of art DCGAN modules, qualitative improvements can be obtained. Blue arrows in Figure~\ref{fig:fig2} point to anatomy missed using the prior GANs investigated. However, we can see that by incorporating our proposed module liver tissue characterization in the synthesized images improves.

\noindent\textbf{Quantitative Results:} The \textit{Inception Score} (IS) and \textit{Frechet Inception Distance} (FID) score are used to quantitatively evaluate the generated image quality and diversity. The \textit{Inception Score} (IS) helps to estimate the quality of the generated images based on the classification performance of Inception V3 classifier on the synthesized images\cite{DBLP:journals/corr/HeuselRUNKH17}. 
Higher IS value means the synthesized images are diverse and similar to the real data \cite{kazeminia2020gans,yi2019generative}. Although IS is a very good metric to assess the quality of the synthesized images, it does not compare the synthetic images with the original images. \textit{Frechet Inception Distance} (FID) is based on the statistics of the generated images compared with that of the original images\cite{DBLP:journals/corr/HeuselRUNKH17}. Similar to IS, FID is also calculated using the Inception V3 model, the activations of the last pooling layer are summarized as a multi-variate Gaussian, the distance between the two Gaussians are calculated as FID\cite{DBLP:journals/corr/HeuselRUNKH17}. A low FID shows that the images synthesized by this GAN architecture have high diversity in them and are at par with the real images\cite{kazeminia2020gans,yi2019generative}. 

The IS and FID metrics are calculated on 400 synthesized images for each category. From Table~\ref{tab:GANresults}, it is noted that our Stage-\uppercase\expandafter{\romannumeral2} GAN module significantly improved the IS and FID results for all the investigated prior GAN modules (paired t-test $p<0.05$). The highest IS score is obtained when DCGAN is combined with the SN and our proposed Stage-\uppercase\expandafter{\romannumeral2} GAN module. The lowest FID score is achieved when DCGAN is combined with the proposed model, that is without the SA and SN modules.  
\begin{table}[]

	\caption{IS and FID of the proposed DCGAN, DCGAN+SN, DCGAN+SN+SA to synthesize liver US images directly and incorporating our Stage-\uppercase\expandafter{\romannumeral2} GAN. Bold text shows the best results obtained. SA- Self attention module, SN- Spectral Normalization module. In all the presented results \textit{ours} denotes the integration of Stage-\uppercase\expandafter{\romannumeral2} GAN module.}
		\label{tab:GANresults}
		\centering \scalebox{0.7}{
		\begin{tabular}{c|c|c}
		\hline
		& \begin{tabular}[c]{@{}c@{}}IS$\uparrow$ \\ abnormal/normal\end{tabular} & \begin{tabular}[c]{@{}c@{}}FID $\downarrow$ \\ abnormal/normal\end{tabular} \\ \hline
		DCGAN\cite{radford2015unsupervised} & 1.32$\pm$0.02 / 1.28$\pm$0.01 & 113.87/161.76 \\ \hline
		DCGAN\cite{radford2015unsupervised}+ours & 1.55$\pm$0.08 / 1.48$\pm$0.05 & \textbf{100.05/99.53} \\ \hline
		DCGAN\cite{radford2015unsupervised}+SN\cite{miyato2018spectral} & 1.09$\pm$0.01 / 1.34$\pm$0.06 & 170.68/247.76 \\ \hline
		DCGAN\cite{radford2015unsupervised}+SN\cite{miyato2018spectral}+ours & \textbf{1.67$\pm$0.08} / 1.50$\pm$0.05 & 156.55/110.17 \\ \hline
		DCGAN\cite{radford2015unsupervised}+SN\cite{miyato2018spectral}+SA\cite{zhang2019self} & 1.42$\pm$0.03 / 1.38$\pm$0.03 & 160.19/259.19 \\ \hline
		DCGAN\cite{radford2015unsupervised}+SN\cite{miyato2018spectral}+SA\cite{zhang2019self}+ours & 1.51$\pm$0.07 / \textbf{1.51$\pm$0.06} & 108.39/103.07 \\ \hline
	\end{tabular}}

\end{table}

\subsection{Classification Evaluation}
To evaluate the quality of the synthesized images, EfficientNet \cite{tan2019efficientnet} is employed to perform binary classification on the original dataset and expanded class-balanced dataset. The training dataset is expanded from 410 images to 1000 images using 590 generated images (Diseased class: 310 real + 190 synthetic; Healthy class: 100 real + 400 synthetic). As explained previously, test data was 140 real US images (70 health 70 diseased) which were not part of the image synthesis process. The classification performance is measured by $accuracy$, $precision$, $recall$ and $F1_{score}$. The quantitative results are shown in Table~\ref{tab:effnetresults}. From the table, it can be noted that the classification algorithm obtains the best accuracy when the synthesized images are obtained using DCGAN in combination with the proposed Stage-\uppercase\expandafter{\romannumeral2} GAN module (paired t-test $p<0.05$). Similar to GAN evaluation results, from Table~\ref{tab:effnetresults} we can observe that our proposed Stage-\uppercase\expandafter{\romannumeral2} GAN module significantly improves classification performance metrics for all the investigated prior GAN modules (paired t-test $p<0.05$).

\begin{table}[]
		\caption{ Quantitative classification results for all the investigated methods. Bold text shows the best results obtained. In all the presented results \textit{ours} denotes the integration of Stage-\uppercase\expandafter{\romannumeral2} GAN module.}
			 
	\centering \scalebox{0.7}{\begin{tabular}{c|c|c|c|c}
	
		\hline
		& Accuracy & \begin{tabular}[c]{@{}c@{}}Precision\end{tabular} & \begin{tabular}[c]{@{}c@{}}Recall\end{tabular} & \begin{tabular}[c]{@{}c@{}}$F1_{score}$\end{tabular} \\ \hline
		The original dataset & 82.14\% & 82.47\% & 82.14\% & 82.10\% \\ \hline
		DCGAN\cite{radford2015unsupervised} & 84.29\% & 84.31\% & 84.29\% & 84.29\% \\ \hline
		DCGAN\cite{radford2015unsupervised}+ ours & \textbf{85.71}\% & \textbf{87.68}\% &\textbf{85.71}\% & \textbf{85.53}\% \\ \hline
		DCGAN\cite{radford2015unsupervised} + SN\cite{miyato2018spectral} & 74.29\% & 76.00\% & 74.29\% & 73.85\% \\ \hline
		DCGAN\cite{radford2015unsupervised} + SN\cite{miyato2018spectral} + ours & 78.57\% & 79.17\% & 78.57\% & 	78.46\% \\ \hline
		DCGAN\cite{radford2015unsupervised} + SN\cite{miyato2018spectral} + SA\cite{miyato2018spectral} & 80.71\% & 80.77\% & 80.71\% & 80.71\% \\ \hline
		DCGAN\cite{radford2015unsupervised} + SN\cite{miyato2018spectral} + SA\cite{miyato2018spectral}   + ours & 82.86\% & 82.88\% & 82.86\% & 82.85\% \\ \hline

	\end{tabular}}

	\label{tab:effnetresults}
\end{table}

\section{Conclusion}

In this work, a novel GAN architecture for realistic B-mode liver US image generation was proposed. Qualitative and quantitative results show significant improvements in image synthesis can be achieved using the proposed two-stage architecture. We also show that the classification performance of well-known CNN architectures can be significantly improved using the synthesized images. Our study is the first attempt to synthesize diseased and healthy liver US images based on a novel GAN module that can be incorporated into popular GAN-based models for improving their performance. One major drawback of our work is the limited dataset size. We only had access to 550 B-mode US data. Increasing the dataset size could also result in the performance improvements of the classification method investigated in this work. Furthermore, we have only evaluated the performance of DCGAN as Stage-\uppercase\expandafter{\romannumeral1} GAN architecture. Investigation of various other GAN architectures, used for medical image synthesis \cite{kazeminia2020gans,yi2019generative}, should also be performed to understand the full potential of our Stage-\uppercase\expandafter{\romannumeral2} GAN model. Finally, a comparison study against traditional augmentation methods should also be performed. Future work will include the collection of large-scale liver US data and improvements of the shortcomings of our work.

\newpage
\bibliographystyle{splncs04}
\bibliography{samplepaper_v2}

\end{document}